\documentclass[aps,prb,twocolumn,showpacs,groupedaddress]{revtex4}
\usepackage{graphicx}
\usepackage{bm}
\begin{document}
\title{
Many-player entangled state solutions in game theory problems}
\author{Sudhakar Yarlagadda}
\affiliation{CAMCS and TCMP, Saha Institute of Nuclear Physics, Calcutta, India}
\date{\today}
\begin{abstract}
{We propose a non-classical multi-player entangled state which eliminates
the need for communication, yet can solve problems (that require
coordination) better than classical approaches.
For the entangled state, 
we propose a slater determinant of all allowed states of a filled band
in a condensed matter system --
 the integer quantum Hall state at filling factor 1.
Such a state 
 gives the best solution (i.e., best Nash equilibrium)
for  some classical stochastic problems where classical solutions
 are far from ideal.}
\end{abstract}

\maketitle

\section{INTRODUCTION}

Game theoretical problems dealing with conflict of interest have been
tackled in the recent past with quantum approaches \cite{landsburg}.
 It is hoped that
quantum game theory, by exploiting quantum mechanics, would
produce significantly improved solutions.
Of particular interest is how many-particle quantum entanglement
can be harnessed to provide better strategies (in games)
compared to the classical solutions.
Entanglement, which  provides correlations between
remote particles, can  equip the players with a coordinated
set of actions depending on the state of the particle that they
privately observe.
Thus when players cannot communicate by classical channels, they can still
arrive at an optimal strategy.

Quantum non-locality was first demonstrated by Bell with his famous
inequalities \cite{bell}. Quantum theory predicts correlations among
outcomes of distant measurements which cannot be explained using only
local variables. It has been demonstrated that two photons are correlated
over large distances (of the order of 10 km) thereby violating Bell's
inequalities \cite{tittel}. Thus we have a verification of the basic
assumption of quantum information and computation that quantum systems can be
entangled over large distances and times.

In the past quantum entanglement has been incorporated
 in classical two-party games such as 
 the prisoner's dilemma by Eisert {\em et al.} \cite{eisert}, the battle of
sexes by Marinatto and Weber \cite{weber}, etc. These authors demonstrated
how optimal solutions can be achieved using entanglement. 
The purpose of the present work is to propose many-particle entangled states
and show how they can be used to obtain improved/optimal solutions for 
classical problems requiring coordinated action by the players.

\section{Many-particle entangled state}
In condensed matter  systems
one frequently encounters bands filled with fermions.
 Based on Pauli's exclusion principle, 
the ground state of any $N$-state
band filled completely by $N$ spinless-fermions is
a Slater determinant of the complete set of
$N$ single particle eigen states of the band.
 Such a Slater determinant
is an antisymmetric linear superposition of $N!$-many $N$-particle
eigen states.
Here {\em we consider a case of a degenerate filled band -- the integer quantum Hall state at
 filling factor 1}.

In our quantum Hall state,
electrons are chosen to be confined to the xy-plane and subjected to
a perpendicular magnetic field. On choosing a symmetric gauge vector
potential, $\vec{A} = 0.5 B (y \widehat {x} -x \widehat{y})$,
 the degenerate single-particle wavefunctions
for the lowest Landau level (LLL) are given by:
\begin{eqnarray}
\phi_m (z) \equiv |m \rangle =
\frac{1}{(2 \pi l_0^2 2 ^m m!)^{\frac{1}{2}}} \left ( \frac{z}{l_0} \right )^m
e^{-|z|^2/4l_0^2} ,
\end{eqnarray}
where $z=x-iy$ is the electron position in complex plane, $m$ is the orbital
angular momentum, and $l_0 \equiv \sqrt{\hbar c /e B}$ is the magnetic length.
The area occupied by the electron in state $|m \rangle$ is
\begin{eqnarray}
\langle m | \pi r^2 |m \rangle = 2(m+1)\pi l_0^2 .
\end{eqnarray}
Thus the LLL can accommodate only $N_e$ electrons given by 
\begin{eqnarray}
N_e = (M+1) = \frac{A}{2 \pi l_0^2} ,
\end{eqnarray}
where  $A$ is the area of the
system and
 $M$ is the largest allowed angular momentum for area $A$.
The many-electron system is described by the Hamiltonian
\begin{eqnarray}
H =  &&
\sum_j \frac{1}{2 m_e} \left [ -i \hbar \nabla_j -  \frac{e}{c} 
\vec{A}_j \right ] ^2 + \sum_{j < k} \frac{e^2}{|z_j - z_k |}
\nonumber \\
&&
 +
 g \mu_B \sum_j
 \vec{B} \cdot \vec{s}_j .
\end{eqnarray}
Thus when the LLL (with the lowest Zeeman energy) is completely filled with
 $N_e$ electrons (i.e., when LLL is at filling factor $\nu =1$),
the many-particle wavefunction $\Psi (z_1, z_2, ....,z_{N_e})$
is given by the Slater determinant 
\begin{eqnarray}
\left| \begin{array}{cccc}
\phi_0(z_1) & \phi_0 (z_2) & \ldots & \phi_0 (z_{N_e}) \\
\phi_1(z_1) & \phi_1 (z_2) & \ldots & \phi_1 (z_{N_e}) \\
\vdots & \vdots & \vdots & \vdots \\
\phi_{N_e -1}(z_1) & \phi_{N_e -1} (z_2) & \ldots & \phi_{N_e -1} (z_{N_e}) 
\end{array} \right| .
\end{eqnarray}
The 
 many-particle wavefunction $\Psi (z_1, z_2, ....,z_{N_e})$
 for $N_e$ particles 
can be
expressed as follows:
\begin{eqnarray}
 \Psi (z_1, z_2, ....,z_{N_e}) = 
 \psi (z_1, z_2, ....,z_{N_e})  
e^{-\sum_{l=1}^{N_e}|z_l|^2/4l_0^2} ,
\label{wf1}
\end{eqnarray}
where
\begin{eqnarray}
\!\!\!\!\!\!
 \psi (z_1, z_2, ....,z_{N_e}) =
&&
\!\!\!\!\!\!
 \prod_{1 \le j < k \le N_e} (z_j -z_k )
\nonumber \\
=
&&
\!\!\!\!\!\!
 \sum_{\sigma \in S_{N_e}} {\rm sgn}(\sigma ) z_1^{\sigma(1)-1} ... 
z_{N_e}^{\sigma(N_e)-1} ,
\label{wf2}
\end{eqnarray}
where $S_{N_e}$ denotes the set of permutations of $\{1,2, ..., N_e \}$
and ${\rm sgn} (\sigma)$ denotes the signature of the permutation $\sigma$.
 Thus we see that $\psi (z_1, z_2, ....,z_{N_e}) $ is a linear
superposition of $N_e!$ states (all with the same probability
of being observed) 
and each state
 $ z_1^{\sigma(1)-1} ...  z_{N_e}^{\sigma(N_e)-1} $
 has the angular momenta $0, 1, 2, ..., N_e -1$ distributed among
$N_e$ fermionic particles (at positions $z_1, z_2, ..., z_{N_e}$) in a uniquely
 different way (with no two particles having the same angular momentum)!
 Thus if the 
 many-particle wavefunction $\Psi (z_1, z_2, ....,z_{N_e})$
is measured for angular momentum of its particles (using for instance
a Stern-Gerlach type of set-up) at positions 
 $z_1, z_2, ..., z_{N_e}$,  then  one of the $N_e!$ permutations
of the angular momentum from the set $\{ 0, 1, 2, ..., N_e -1 \}$
will be measured with probability $1/(N_e !)$.
The above fact can be exploited in a game-theoretic context as described
in the next section.

Here it should be pointed out
that although an antisymmetric wavefunction obtained based
on Pauli's exclusion principle is in general not an entangled state \cite{shi,peres},
the Coulomb interactions actually produce the same antisymmetric
wavefunction even when the 
fermionic nature of the particles is ignored,
i.e., for example if the particles are treated as classical particles.
Furthermore, for the situation where the g-factor is zero (which can
be achieved in gallium arsenide heterostructures using pressure),
Coulomb interaction energy is minimized when the real space wave
function is antisymmetric and given by Eq. (\ref{wf1}) while the
spin wavefunction is symmetric (with the total spin being maximized
and equal to $N_{e}/2$). This is clearly an entangled state based on correlation effects.
This situation is very similar to that of the electronic wavefunction
in  a half-filled  degenerate sub-shell in an atom (such as the five electrons
 in the  3d sub-shell of $Mn^{2+}$)  where Hund's rule dictates that
wavefunction be antisymmetric in the real space and symmetric in the spin space.
 In general, for the quantum Hall situation (at filling factor 1) where one
has at least two species of fermionic 
particles  with all the particles having
the same charge, spin, and 
single particle energy ($\hbar \omega_c/2 - 0.5 g \mu_B B$),
one again obtains [for total number of particles
$N= N_e = A/(2 \pi l_0^2$)] the same many-body
wavefunction [given by Eq. (\ref{wf1})]
which now is certainly entangled due to correlation effects
produced by Coulomb interactions.
Lastly, we would like to add that, the above considerations
for minimum Coulomb interaction energy are certainly valid when the 
repulsive interaction
is given by a short range  
Dirac-delta function in which case the interaction energy is zero.

\section{Quantum solutions to classical problems}
In this section we will pose a 
couple of
 classical problems and
show that entanglement not only significantly improves the solution,
in fact, it also produces the best possible solution.

\subsection{Kolkata restaurant problem}
We will first examine the Kolkata restaurant problem (KRP) \cite{bkc1} 
which is a variant of the Minority Game Problem \cite{minority}.
In the KRP  (in its minimal form) there are $N$ restaurants 
(with $N \rightarrow \infty$) that can
each accommodate only one person and there are $N$ agents to be accommodated.
All the $N$ agents take a stochastic strategy that is independent 
of each other.
If we assume that, on any day,
 each of the $N$ agents chooses randomly any of
the $N$ restaurants such that if $m$ ($> 1$) agents show up at any restaurant,
then only one of them (picked randomly) will be served and the rest $m-1$
go without a meal. It is also understood that each agent can choose only
one restaurant and no more. Then the probability $f$ that a person gets
a meal (or a restaurant gets a customer) on any day is calculated based
on the probability $P(m)$ that any restaurant gets chosen by $m$ agents
with
\begin{eqnarray}
 P(m)
=
\frac{N!}{(N-m)! m!}p^m (1-p)^{N-m} 
=
\frac{\exp(-1)}{m!} ,
\end{eqnarray}
where $p=1/N$ is the probability of choosing any restaurant.
Hence, the fraction of restaurants that get chosen on any day
is given by
\begin{eqnarray}
f= 1-P(0) = 1- \exp(-1) \approx 0.63 .
\end{eqnarray}

\begin{figure}
\includegraphics[width=3.0in]{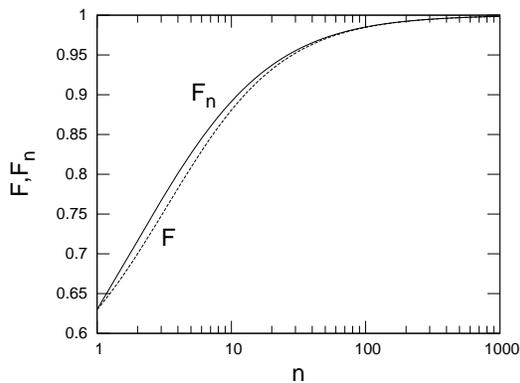}
\noindent\caption[]
{Plot of the exact probability $F_n$
and the continuum approximation probability $F$
 for various iteration values $n$.}
\label{fig1}
\end{figure}

Now, we extend the above minimal KRP game to get a more efficient
utilization of restaurants by taking advantage of past experience
of the diners.  We stipulate that the 
 successful diners ($NF_n$)
 on the $n$th day will visit the 
same  restaurant on all subsequent days as well,
 while the remaining $N-NF_n$ unsuccessful agents of the
$n$th day try stochastically any of the
 $N$ restaurants on the next day (i.e., $n+1$th day)
 and so on until all customers find a restaurant.
The above procedure can be mathematically modeled to yield the
following recurrence relation
\begin{eqnarray}
F_{n+1}=F_{n}+f(1-F_n)^2 ,
\label{F_n}
\end{eqnarray}
 where $F_n$ is the fraction of restaurants
 occupied on the $n$th day with
 $F_1 = f= 1-1/e$.
Upon making a continuum approximation we get
\begin{eqnarray}
\frac{dF}{dn}= f(1-F)^2 ,
\end{eqnarray}
which yields the solution
\begin{eqnarray}
F = 1 - \frac{e}{e^2+(n-1)
 (e-1)} .
\end{eqnarray}
The above solution $F$ turns out to be a good approximation to
the solution for $F_n$ in Eq. (\ref{F_n}) (with error less than 5\%)
as can be seen from Fig. \ref{fig1}.  
We see that even after 10 iterations less than 90\% of the
restaurants are filled!

We will now investigate how superior quantum solutions can be obtained for the
 KRP.
We will introduce quantum mechanics into the problem
 by asking the $N$ agents to share
an entangled $N$-particle quantum Hall state at filling factor 1 described
in the previous section [see Eqs. (\ref{wf1}) \& (\ref{wf2})].
 We assign to each of the $N$ restaurants
a unique angular momentum picked from the set
 $\{ 0, 1, 2, ..., N -1 \}$.  We ask each agent to measure the angular momentum
of a randomly chosen particle from the $N$-particle entangled state.
Then, based on the measured angular momentum, the agent goes to the restaurant
that has his/her particular angular momentum assigned to it. In this
 approach all the agents get to eat in a restaurant and all the restaurants
get a customer. Thus we see that the prescribed entangled state
always produces restaurant-occupation probability 1 
 and is thus superior to the classical solution mentioned above!
Furthermore, the probability that an agent picks a restaurant is still $p=1/N$
and hence
all agents are equally likely to go to any restaurant. Thus, even if there
is an accepted-by-all hierarchy amongst the restaurants (in terms of quality
 of food with price of all restaurants being the same),
 the entangled state  produces an equitable (Pareto optimal) solution where
 all agents
have the same probability of going to the best restaurant, or the
 second-best restaurant, and so on. 
{\it Quite importantly, it can be shown that the chosen
entangled quantum strategy
(i.e., the entangled $N$-particle quantum Hall state at filling factor 1)
 actually represents the best Nash equilibrium when there is a restaurant
ranking!} (see Appendix A for details).

\subsection{Kolkata stadium problem}
We will next introduce a variant of the KRP game which we will call
as the Kolkata stadium problem (KSP). In the KSP, there are $NK$ agents
trapped inside a Kolkata stadium that has $K$ exits.
There is a panic situation of a fire or a bomb-scare and all the
agents have to get out quickly
 through the $K$ exits each of which has a capacity
of $\alpha N$
with $\alpha \ge 1$. We assume that all $NK$ agents have
equal access to all the exits and that each agent has
enough time to approach only one exit
before being harmed.
The probability $P(m)$ that any exit gets chosen by $m$ agents is given by
the binomial distribution
\begin{eqnarray}
 P(m)
=
\frac{(NK)!}{(NK-m)! m!}p^m (1-p)^{NK-m} ,
\end{eqnarray}
where $p=1/K$ is the probability of choosing any gate.
For a capacity of $\alpha N$ for each gate, the cumulative probability
$P=\sum_{m=1}^{\alpha N}P(m)$ that
(on an average) a gate is approached by
  $\alpha N$ or less agents is given in Table \ref{table1}.

\begin{table}
\caption{\label{table1} The calculated values of the cumulated probability
$P$ for a system with $NK$ persons and $K$ gates with a gate-capacity
 $\alpha N$.}
\begin{ruledtabular}
\begin{tabular}{|c|c|c|c||c||c|c|c|c|}
\hline
  \multicolumn{1}{|c|}{{\bf $\alpha$}}&
   \multicolumn{1}{c|}{{\bf $N$}}&
\multicolumn{1}{c|}{{\bf $K$}} &
\multicolumn{1}{c||}{{\bf $P$}} &
\multicolumn{1}{c||}{{\bf }} &
  \multicolumn{1}{c|}{{\bf $\alpha $}} &
   \multicolumn{1}{c|}{{\bf $N$}}&
\multicolumn{1}{c|}{{\bf $K$}} &
\multicolumn{1}{c|}{{\bf $P$}}
\\
\hline
1   & 100  & 10  & 0.5266  && 1.05   & 100  & 10  & 0.7221  \\
1   & 500  & 10  & 0.5119  && 1.05   & 500  & 10  & 0.8848  \\
1   & 1000  & 10  & 0.5084 & & 1.05   & 1000  & 10  & 0.9531  \\
1   & 10000  & 10  & 0.5027 & & 1.05   & 10000  & 10  & 1.0000  \\
\hline
\hline
1   & 100  & 20  & 0.5266  && 1.1   & 100  & 10  & 0.8652  \\
1   & 500  & 20  & 0.5119  && 1.1   & 500  & 10  & 0.9907  \\
1   & 1000  & 20  & 0.5084 & & 1.1   & 1000  & 10  & 0.9995  \\
1   & 10000  & 20  & 0.5027 & & 1.1   & 10000  & 10  & 1.0000  \\
 \hline
\end{tabular}
\end{ruledtabular}
\end{table}

Thus we see that if a gate has the optimal capacity of $N$ (i.e., $\alpha = 1$),
then $P$ is close to $0.5$ and is not affected by the number of gates $K$
(for small $K$) with $ P \rightarrow 0.5$ for $N\rightarrow \infty$.
However, as $\alpha$ increases even slightly above unity, $P$
increases significantly for fixed values $N$ and $K$.
Furthermore, for fixed values of $\alpha >1$ and $K$ (with $\alpha$ only
slightly larger than 1 and $K$ being small) $P \rightarrow 1$ as $N$ becomes
large.
{\em Here it should be mentioned that
even when $P \rightarrow 1$ on an average,
 there can be fluctuations in a stampede
situation with more than $\alpha N$ people approaching a gate
and thus resulting in fatalities}.

Here too in the KSP game, if the $NK$ agents were to use the entangled
$NK$-particle state given by
the quantum Hall effect state at filling factor 1,
 then every agent is assured of safe passage. In this situation, since there
are $NK$ angular momenta and only $K$ gates,
 the angular momentum $M_i$ measured by an agent $i$ for
his/her particle should be divided
by $K$ and the remainder be taken to give the appropriate gate number
[i.e., gate number = $M_i$ ({\bf mod} $K$)]. Thus entanglement gives safe exit
with probability 1 even when $\alpha =1$!
Furthermore, even if there is an accepted-by-all ranking of the exits
in terms of comfort of passage, our chosen entangled state corresponds
to the best Nash equilibrium!

\section{Conclusions}
In the $N$-agent KRP game, while the number of  satisfactory choices
is only $N!$, in sharp contrast the number of possibilities is $N^N$ when
all the restaurants have the same ranking.
Thus, in the classical stochastic approach,
 the probability of getting the best solution where all the restaurants are
occupied by one customer is given by the vanishingly small value $\exp(-N)$.
Even in the KSP case, it can be shown that there is a vanishingly
small probability [=$\sqrt{K/(2\pi N)^{K-1}}$] of providing safe
passage to all when only $N$ people are allowed to exit from each of
the $K$ gates (i.e., when $\alpha=1$).
On the other hand,
in this work we showed how quantum entanglement can produce a coordinated
action amongst all the $N$-agents leading to the best possible solution
with a probability 1!.
Thus quantum entanglement produces a much more desirable scenario
compared to a classical approach at least for the KRP  and the KSP games.

As a candidate for entanglement we could have picked any filled
band system (of condensed matter physics)
 even in the absence of a magnetic field. For such an
entangled $N$-particle
state, momentum is a good quantum number.
 However, only when the Coulomb interaction
is infinity compared to the kinetic energy do we have the
many-body ground state given by the antisymmetric wavefunction
 satisfying Pauli's exclusion
principle.
Furthermore, the minimum
spacing between
various particle momenta is only $2\pi \hbar /L$ where $L$ is the linear size
of the system
 and hence, to unambiguously determine the momentum of a particle, 
 one is faced with the uncertainty principle
which fixes the uncertainty in the measured  momentum 
 to be at least $\hbar/2L$.

 Next, one can also consider
$N$ number of identical qudits
each with $N$ possible states.
 By  producing
an antisymmetric entangled state from these N qudits,
one can get better results than  classical approaches.
However, physically realizing a 
qudit with a large number of states
is a challenging task \cite{qudit}.

Lastly, although it has not been shown that 
our many-particle entangled state
(i.e., the quantum Hall effect state at filling factor 1)
 will have 
long-distance and also long-term correlations, we are
hopeful of such a demonstration in the future.

\section{ACKNOWLEDGEMENTS}
The author would like to thank Bikas K. Chakrabarti and  K. Sengupta
for useful discussions.
Furthermore, discussions with R. K. Monu
on the literature are also gratefully acknowledged.

\appendix
\section{}
In a $N$-player game, the set of strategies
$(s_1^* , s_2^* , ..., s_N^* )$ represent a {\it Nash Equilibrium}
if, for every player $i$, the strategy $s_i^*$ meets the following requirement
for the payoff function $\$$:
\begin{eqnarray}
 &&
\$_i(s_1^*, ....,s_{i-1}^* , s_i^* , s_{i+1}^* , ....,s_N^*)
\nonumber
\\
&&  ~~
\ge 
\$_i(s_1^*, ....,s_{i-1}^* , s_i , s_{i+1}^* , ....,s_N^*) ,
\end{eqnarray}
for every strategy $s_i$  available to $i$.
In order to illustrate the main idea behind exploiting quantum
strategies, we will consider the simple situation of
two restaurants $R_1$ and $R_2$ with utility $u_1$ and $u_2$ respectively
as perceived by two diners $D_1$ and $D_2$. Then we can represent the payoff
for the diners by using the  bimatrix displayed in
 Table \ref{table2} with diner $D_1$ choices
 along the rows and those of $D_2$ along the columns.
\begin{table}
\caption{\label{table2}}
\begin{tabular}{|c|c|c|}
  \multicolumn{1}{c|}{{}}&
  \multicolumn{1}{c|}{{\bf $R_1$}}&
  \multicolumn{1}{c|}{{\bf $R_2$}}
\\
\hline
  \multicolumn{1}{c}{{$R_1$}}&
  \multicolumn{1}{|c|}{{\bf $\left ( \frac{u_1}{2},\frac{u_1}{2} \right ) $}}&
   \multicolumn{1}{c|}{{\bf ( $u_1$,$u_2$ )}}
\\
\hline
  \multicolumn{1}{c}{{$R_2$}}&
   \multicolumn{1}{|c|}{{\bf ( $u_2$,$u_1$ )}} &
  \multicolumn{1}{c|}{{\bf $ \left ( \frac{u_2}{2},\frac{u_2}{2} \right )$}}
\\
\hline
\end{tabular}
\end{table}

Here we use the formalism developed 
 in Ref. \onlinecite{weber}.
We assume that diners $D_{1,2}$ have access to the following entangled state:
\begin{eqnarray}
|\psi_{in} \rangle = a|R_1 R_2 \rangle + b | R_2 R_1 \rangle ,
\label{psi_in}
\end{eqnarray}
where the coefficients satisfy the condition
$|a|^2 + |b|^2 =1$.
The corresponding initial density matrix is given by
\begin{eqnarray}
\rho_{in} = \rho_{in} ^{D_1} \otimes \rho_{in} ^{D_2} =
|\psi_{in} \rangle \langle \psi_{in} | .
\end{eqnarray}
We assume that each player can manipulate his state (i.e., restaurant) by
either using the identity $I$ or the Pauli flipping operator $\sigma_x$
which is unitary and has the following property
\begin{eqnarray}
\sigma_x |R_{1,2} \rangle = | R_{2,1} \rangle .
\end{eqnarray}
We further assume that each diner can transform his part
 ($\rho_{in} ^{D_{1,2}}$)
of the total density matrix $\rho_{in} $ in the following manner:
\begin{eqnarray}
\rho_{fin}^{D_{1,2}} = p_{1,2} I \rho_{in} ^{D_{1,2}} I^{\dagger}
+
(1- p_{1,2}) \sigma_x \rho_{in} ^{D_{1,2}} \sigma_x^{\dagger} ,
\end{eqnarray}
to obtain the final density matrix
\begin{eqnarray}
\rho_{fin} = \rho_{fin}^{D_{1}} \otimes \rho_{fin}^{D_{2}}  .
\end{eqnarray}
In order to evaluate the payoff, we define the payoff operator as follows:
\begin{eqnarray}
\!\!\!\!\!\!\!\!\!\!
P_{1,2} = &&
\!\!\!\!\!\!
u_{1,2}|R_1 R_2 \rangle \langle R_1 R_2 |
+ u_{2,1} |R_2 R_1 \rangle \langle R_2 R_1 |
\nonumber 
\\
&& 
\!\!\!\!\!\!\!
+ 0.5 u_1 |R_1 R_1 \rangle \langle R_1 R_1 |
+ 0.5 u_2 |R_2 R_2 \rangle \langle R_2 R_2 | .
\end{eqnarray}
Then the payoffs obtained using the following expression
\begin{eqnarray}
\$_{1,2} = {\rm Tr} (P_{1,2} \rho_{fin}) ,
\end{eqnarray}
are given by
\begin{eqnarray}
\$_{1}(p_1,p_2) = &&
\!\!\!\!\!\!
0.5 p_1 p_2 (u_1+u_2) 
\nonumber
\\
&&
\!\!\!\!\!\!
+ p_1 \left [ 0.5 u_1 |a|^2 + 0.5 u_2 |b|^2 - u_2 |a|^2 - u_1 |b|^2 \right ]
\nonumber
\\
&&
\!\!\!\!\!\!
- 0.5  p_2 \left [  u_1 |b|^2 +  u_2 |a|^2 \right ]
\nonumber
\\
&&
\!\!\!\!\!\!
+ u_2 |a|^2 + u_1 |b|^2 ,
\end{eqnarray}
and
\begin{eqnarray}
\$_{2}(p_1 , p_2 ) = &&
\!\!\!\!\!\!
0.5 p_1 p_2  (u_1+u_2)
\nonumber
\\
&&
\!\!\!\!\!\!
- 0.5 p_1 \left [ u_1 |a|^2 + u_2 |b|^2 \right ]
\nonumber
\\
&&
\!\!\!\!\!\!
+ p_2 \left [ 0.5 u_1 |b|^2 + 0.5 u_2 |a|^2 - u_1 |a|^2 - u_2 |b|^2 \right ]
\nonumber
\\
&&
\!\!\!\!\!\!
+ u_1 |a|^2 + u_2 |b|^2 .
\end{eqnarray}
To determine the Nash equilibria, we stipulate that the following
differences be non-negative:
\begin{eqnarray}
\$_{1}(p_1^*,p_2^*)
- \$_{1}(p_1,p_2^*)
 = 
(p_1^* -p_1)
&&
\!\!\!\!\!\!
 \left [ 
 0.5 p_2^* (u_1+u_2) 
\right .
\nonumber
\\
&&
\!\!\!\!\!\!
+ u_1 (0.5 |a|^2 - |b|^2 )
\nonumber
\\
&&
\!\!\!\!\!\!
\left . 
- u_2 (|a|^2 - 0.5 |b|^2) \right ],
\nonumber
\\
\label{diff1}
\end{eqnarray}
and
\begin{eqnarray}
\$_{2}(p_1^*,p_2^*)
- \$_{2}(p_1,p_2^*)
 = 
(p_2^* -p_2)
&&
\!\!\!\!\!\!
 \left [ 
 0.5 p_1^* (u_1+u_2) 
\right .
\nonumber
\\
&&
\!\!\!\!\!\!
+ u_2 (0.5 |a|^2 - |b|^2 )
\nonumber
\\
&&
\!\!\!\!\!\!
\left . 
- u_1 (|a|^2 - 0.5 |b|^2) \right ].
\nonumber
\\
\label{diff2}
\end{eqnarray}
Then, from Eqs. (\ref{diff1}) and (\ref{diff2}),
 we obtain the three Nash equilibria $(p_1,p_2) = (1,1), (0,0)$,
and $(\bar{p}_1, \bar{p}_2)$
where
\begin{eqnarray}
\!\!\!\!\!
\bar{p}_1 \equiv
  - [u_1 (1-3|a|^2)+u_2 (-2+3|a|^2)]/(u_1 +u_2) ,
\end{eqnarray}
and 
\begin{eqnarray}
\!\!\!\!\!
\bar{p}_2 \equiv
 - [u_1 (-2+3|a|^2)+u_2 (1-3|a|^2)]/(u_1 +u_2) .
\end{eqnarray}
Next, we note that the differences
\begin{eqnarray}
\$_{1}(1,1)
- \$_{1}(0,0) = (u_2 - u_1)(1 - 2 |a|^2 ) ,
\end{eqnarray}
and
\begin{eqnarray}
\$_{2}(1,1)
- \$_{2}(0,0) = -(u_2 - u_1)(1 - 2 |a|^2 ) ,
\end{eqnarray}
are equal in magnitude but opposite in sign. Hence to obtain the same
preferred
Nash equilibrium [among (1,1) and (0,0)] for both the diners $D_1$ and $D_2$,
we take $|a| =1/\sqrt{2}$ which makes the payoff (for both the diners)
the same at both the
equilibrium points, i.e., 
$\$_{1,2}(1,1)
= \$_{1,2}(0,0) = (u_2 + u_1)/2$.
It can easily be verified, for the third Nash equilibrium
strategy, that
$\$_1(\bar{p}_1,\bar{p}_2) =
\$_2(\bar{p}_1,\bar{p}_2) \le 3(u_1 + u_2)/8 < (u_1 +u_2)/2$.
Thus the entangled state 
\begin{eqnarray}
|\psi_{in} \rangle = \frac{|R_1 R_2 \rangle -  | R_2 R_1 \rangle}{\sqrt{2}} ,
\end{eqnarray}
corresponds to the best Nash equilibrium.
It can also be argued from the symmetry of the payoff
for the two diners, as shown in Table \ref{table2}, that
one expects the best Nash equilibrium to occur when $|a|=1/\sqrt{2}$
in Eq. (\ref{psi_in}).
The above argument can be extended to the case of
 $N$ diners and $N$ restaurants
each with a different ranking \cite{sudhakar}
and one can deduce that the best Nash equilibrium strategy corresponds
to the many-particle entangled state (i.e., the integer quantum Hall
state at filling factor 1) chosen by us.

\end{document}